\documentclass{aa}
\usepackage{graphicx}
\usepackage{txfonts}

\begin{document}
\title{OGLE-2002-BLG-360: \\from a gravitational microlensing candidate to\\
    an overlooked red transient 
 \thanks{Based on observations obtained with the 1.3-m Warsaw telescope at
the Las Campanas Observatory of the Carnegie Institution for Science.}
 \thanks{ We dedicate
this paper to the memory of the late Professor Bohdan Paczy\'nski who,  
from its discovery, traced the strange  
behaviour of OGLE-2002-BLG-360 with great interest. 
The analysis of the light curve of this
unusual object maintained his excitement for science, despite
serious illness.}}
\author{R. Tylenda\inst{1}
\and T. Kami{\'n}ski\inst{6,1}
\and A. Udalski\inst{2,3}
\and I. Soszy{\'n}ski\inst{2,3}
\and R.~Poleski\inst{2,3,7}
\and M. K. Szyma{\'n}ski\inst{2,3}
\and M.~Kubiak\inst{2,3} 
\and G.~Pietrzy{\'n}ski\inst{2,3,4}
\and S. Koz\l{}owski\inst{2,3}
\and P. Pietrukowicz\inst{2,3}
\and K. Ulaczyk\inst{2,3}
\and {\L}. Wyrzykowski\inst{5,3} 
}

\offprints{R. Tylenda}
\institute{Department for Astrophysics, N.~Copernicus
            Astronomical Center, Rabia\'nska~8, 87-100~Toru\'n, Poland\\
            \email{tylenda@ncac.torun.pl}
     \and Warsaw University Observatory, Al.~Ujazdowskie~4, 
          00-478~Warsaw, Poland\\ 
     \email{udalski,soszynsk,rpoleski,msz,mk,pietrzyn,simkoz,pietruk,kulaczyk@astrouw.edu.pl}
    \and  The Optical Gravitational Lensing Experiment
    \and Universidad de Concepci{\'o}n, Departamento de Astronomia,
         Casilla 160--C, Concepci{\'o}n, Chile
    \and  Institute of Astronomy, University of Cambridge, Madingley Road,
          Cambridge CB3~0HA,~UK \\
          \email{wyrzykow@ast.cam.ac.uk}
    \and Max-Planck-Institut f{\"u}r Radioastronomie, Auf dem H{\"u}gel 69, 
         53121 Bonn, Germany \\
         \email{kaminski@mpifr.de}
    \and Department of Astronomy, Ohio State University, 140 W. 18th Ave., 
         Columbus, OH 43210, USA
}

\date{Received; accepted}

\abstract
{OGLE-2002-BLG-360 was discovered as a microlensing candidate by the
OGLE-III project. The subsequent light curve, however, clearly
showed that the brightening of the object could not have resulted 
from the gravitational microlensing phenomenon.}
{We aim to explain the nature of OGLE-2002-BLG-360 and its eruption
observed in 2002--2006.}
{The observational data primarily come from the archives of the OGLE
project, which monitored the object in 2001--2009. The archives of the
MACHO and MOA projects also provided us with additional data obtained in 1995--99
and 2000--2005, respectively.
These data allowed us to analyse the light curve of the object
during its eruption, as well as the potential variability of its progenitor. In
the archives of several infrared surveys, namely 2MASS, MSX, {\it Spitzer},
AKARI, WISE,
and VVV, we found measurements of the object, which allowed us to study the
spectral energy distribution (SED) of the object.
We constructed a simple model of a star surrounded by a dusty
envelope, which was used to interpret the observed SED.}
{Our analysis of the data 
clearly shows that OGLE-2002-BLG-360 was most probably a red transient,
i.e. an object similar in nature to V838~Mon, whose eruption was observed in
2002. The SED in all phases, i.e. progenitor, eruption, and remnant,
was dominated by infrared emission, which we interpret as evidence of dust
formation in an intense mass outflow. Since 2009 the object has been completely
embedded in dust.}
{We suggest that
the progenitor of OGLE-2002-BLG-360 was a 
binary, which had entered the common-envelope phase a long time (at
least decades) before the observed eruption, and that the eruption resulted
from the final merger of the binary components.
We point out similarities between OGLE-2002-BLG-360 and CK~Vul,
whose eruption was observed in 1670--72, and this strengthens the hypothesis that
CK~Vul was also a red transient.}


\titlerunning{OGLE-2002-BLG-360: a red transient}
\authorrunning{Tylenda et al.}
\maketitle
\section{Introduction  \label{intro}}

Red transients, also known as V838~Mon-type objects,
form a small and heterogeneous but also interesting and 
exciting class of eruptive objects.
The powerful outburst of V838 Mon in 2002 \citep{mun02,crause03}, 
accompanied by a spectacular light echo \citep{bond03},
raised the interest of astrophysicists in this type of stellar eruption.
From their light curves they can be classified
as slow or unusual novae, but they are neither. The most important
characteristics that significantly differentiate red transients from classical novae 
is that the former ones evolve to low effective temperatures 
during outburst and decline as very cool (super)giants.
All known red transient remnants are now observed as late M-type giants and/or
luminous infrared objects. This allowed \citet{soktyl03} and
\citet{tylsok06} to conclude that the red transients cannot be accounted for by
thermonuclear runaways (classical nova, late He-shell flash) and to propose
that these eruptions result from stellar mergers.

\begin{figure*}
  \includegraphics[height=\hsize]{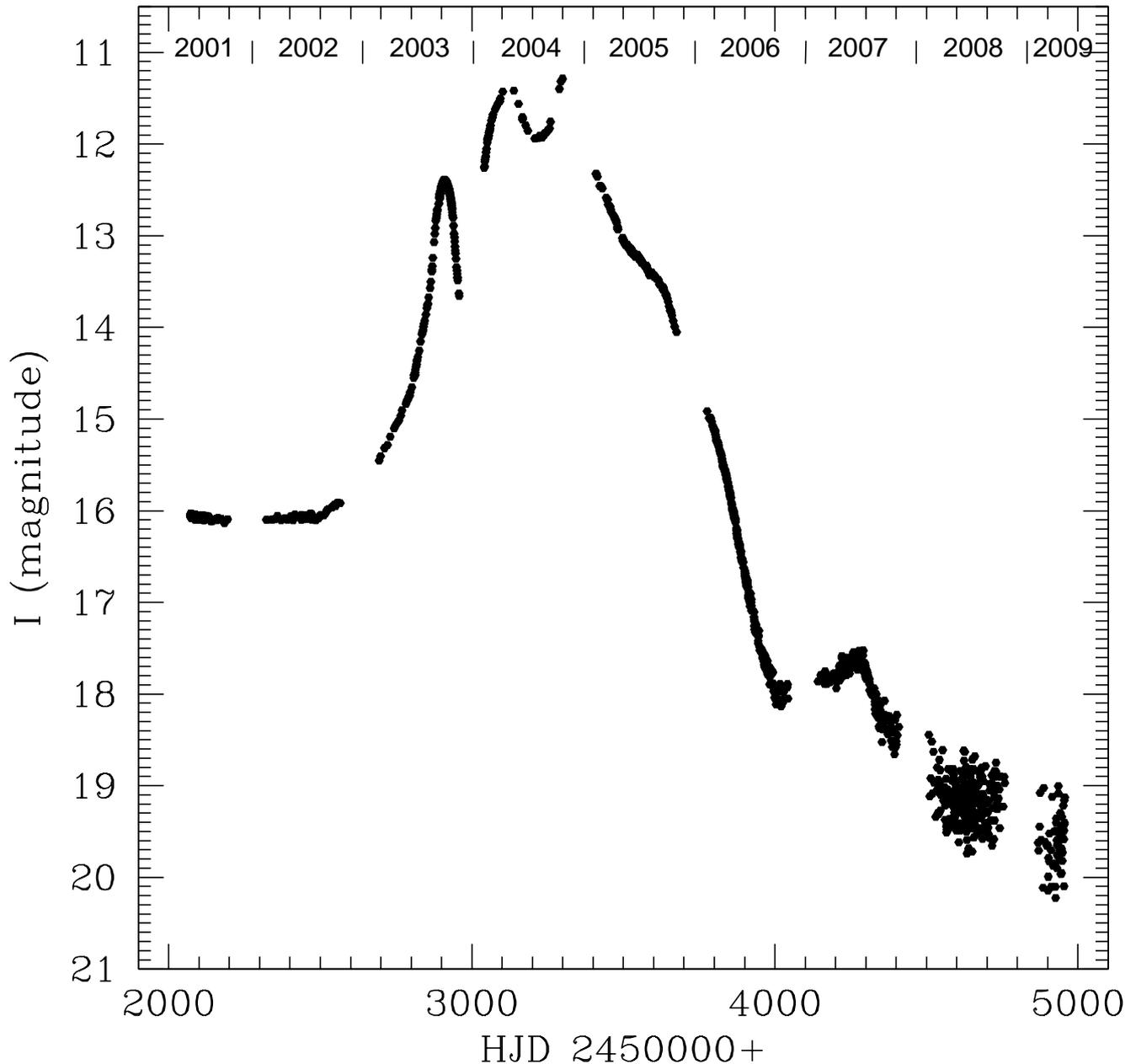}
  \caption{Light curve of OGLE-2002-BLG-360 from the OGLE-III
project: $I$ Kron-Cousins magnitude versus
time of observations in Julian Dates. Time in years is marked on top of
the figure.}
  \label{lc_fig}
\end{figure*}

Apart from V838~Mon, in our Galaxy the class of red transients includes V4332~Sgr, 
whose outburst was observed in 1994 \citep{martini}, and
V1309~Sco, which erupted in 2008 \citep{mason10}. As extragalactic eruptions
of a similar kind one can mention M31~RV \citep[eruption in 1989,][]{mould},
M85~OT2006 \citep{kulk07}, NGC300~OT2008 \citep{bond09,berger09}, and SN~2008S
\citep{smith09}.

V1309 Sco appeared to be a clue object in understanding the nature of red
transients. An analysis of archive photometric measurements, available from the
OGLE project, allowed \citet{thk11} to show that the progenitor of V1309~Sco
was a contact binary system quickly evolving to its merger.

This finding does not necessarily imply that all the above mentioned objects
are stellar mergers. Indeed, it has been seggested that some of the extragalactic
transients, especially those called ''supernova impostors'', 
are due to some other phenomenon involving massive stars
\citep{humph,kochan,smith} or massive binaries \citep{kfs10}, although a
stellar merger scenario cannot be excluded either \citep{ks13}.

In the archives of the OGLE project we have found an object, which
suffered from a several years lasting eruption. Already a superficial analysis of
the data indicates that it might have been an overlooked red transient. The present
paper reports on a detailed analysis of the data on the object, both,
from the OGLE photometric measurements as well as from other available
sources.

\section{The optical light curve  \label{lc_sect}}

The object bears a designation of OGLE-2002-BLG-360 (hereinafter BLG-360 for
short) in the archives of the OGLE project. 
Its coordinates are $\alpha_{2000} = 17^h 57^m 38\fs 97$, $\delta_{2000} =
-29\degr 46\arcmin 04\farcs 8$ ($l = 0\fdg6220$, $b = -2\fdg6788$).

The object was discovered on 9~October 2002 
by the OGLE Early Warning System 
\citep[for a description of the OGLE EWS see][]
{udal03}\footnote{http://ogle.astrouw.edu.pl}. Its early (2002--2003) 
light curve was classified as due a long microlensing event \citep{bep03}. The
subsequent observations, however, ruled out this interpretation.

The OGLE-III project provided us with 2400 measurements in 
the $I$ Kron-Cousins photometric band in 2001--2009. 
The data were reduced and calibrated using standard OGLE procedures
\citep{udal08}. A typical precision of the measurements was
0.01~mag, when the object was brighter than $I \simeq 17$. 

All the OGLE data are displayed in Fig.~\ref{lc_fig}.
The gaps in the data are owing to conjunctions
of the object with the Sun. Apart from 2001 (start of OGLE-III) 
and 2009 (transition period between OGLE-III to OGLE-IV) most of the data were
obtained between February and October of each year. 

As can be seen from Fig.~\ref{lc_fig}, the object was fairly constant in
brightness in 2001 and during first few months of 2002. In August~2002 the
object started a slow brightening, continued with a faster pace in 2003. A
maximum of $I \simeq 12.4$ was reached at the end of September~2003. The
subsequent decline was observed until the end (mid-November) of 
the 2003 observing season. The observations resumed in first days of
February~2004 revealed, however, the object to be again rising in
brightness. A second maximum, about 1\,mag brighter than the first
one, was observed in April~2004. It was followed by a 0.5~magnitude
drop and a subsequent resumption of the brightening. A third maximum,
probably even brighter than the second one, presumably occurred in
October-November~2004. The conjunction of the object with the Sun did not
allow us to establish neither the date of the maximum nor the maximum
brightness. A maximum recorded brightness was of $I = 11.29$ and it was 
observed on 21~October~2004. 

The observations resumed in February~2005 showed the object to be fading.
The decline continued the entire 2005 season and most of the 2006 year.
During two years the object dropped in brightness by $\sim$7~mag in 
the $I$ band. In September the object stabilized at $I \simeq 18.0$. In
2007, a slow rise lasting until July~2007 and a subsequent decline formed
a shoulder in the light curve of BLG-360. In 2008 and 2009 the object
continued fading. BLG-360 is situated in a crowded stellar field and has
a $\sim$19\,mag blending field star at an angular distance of $\sim$0\farcs 4.
Depending on the quality of the OGLE images the companion more or less
contaminated the measurements of BLG-360, when the later was of comparable or
lower brightness than that of the companion. This was the main reason of the
scatter of the observational points seen in Fig.~\ref{lc_fig} in 2008. In
2009, visual inspection of the images shows that BLG-360 was practically
invisible and the OGLE measurements refer rather to the scattered light from
the companion.

The phase III of OGLE terminated in May~2009. The OGLE-IV resumed
observations in March~2010. Since that date until the present epoch
no measurable object has been seen at the position of BLG-360. We can
establish a conservative upper limit to the present brightness of BLG-360 
to be  $I \ga 21$.

\section{OGLE-2002-BLG-360 as a red transient  \label{rn_sect}}

\begin{figure}
  \includegraphics[height=\hsize]{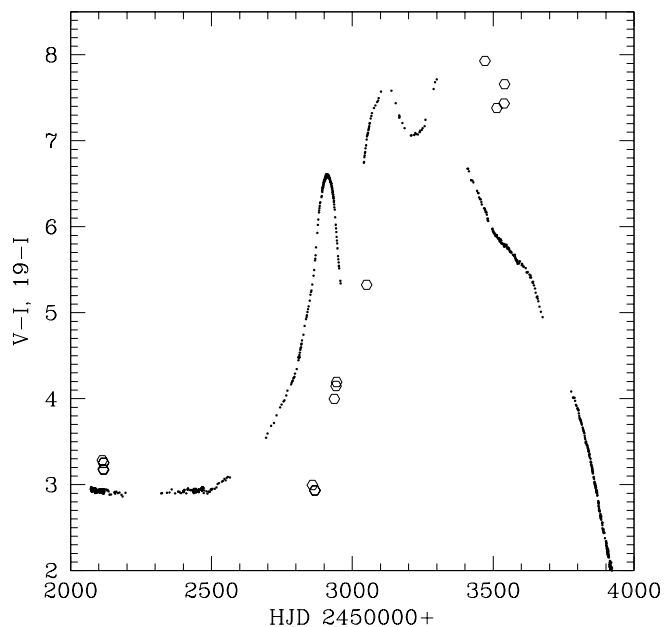}
  \caption{Evolution of the $V - I$ colour of OGLE-2002-BLG-360 -- 
  open symbols. The
  $I$ curve of the object is also shown with small dots.}
  \label{VI_fig}
\end{figure}

Apart from the observations in the $I$ filter, OGLE-III
also provided us with 16 measurements of BLG-360 in the $V$ filter.
This allowed us to study the evolution of the $V - I$ colour during the
eruption of the object. The results are displayed in Fig.~\ref{VI_fig}.

In 2001, i.e. before the eruption, the object had a colour 
$V - I = 3.23 \pm 0.05$. When interpreted with standard colours of giants
(luminosity class III) with no interstellar reddening, 
it would imply a spectral type of M\,5--6. If the star were reddened with 
$E_{B-V} \simeq 1.0$ (see below) then the spectral type would be $\sim$\,M\,1.

During the initial rise in August~2003, the colour was slightly bluer, i.e.
$V - I \simeq 2.95$. However, at the end of October~2003, when the object
was declining after the first maximum, the colour became
significantly redder, i.e. $V - I \simeq 4.1$. The reddening tendency was 
continued in course of the eruption with $V - I$ being 5.3 in February~2004
(rise to the second maximum),
and 7.3 -- 7.9 in April -- June~2005 when the object was declining from the
eruption. The latter colour implies a spectral type of M\,9 or later,
almost independently of the assumed interstellar extinction. The real colour
of the object in 2005 was probably even redder as the $V$ measurements at a
level of 20.5--21.0\,mag were likely to be contaminated by faint field stars
in the crowded field of BLG-360. 

The progressive reddening of BLG-360 during
the eruption and decline is also confirmed by a comparison of the archive
photometric data from the MOA project to the OGLE-III photometry, as done in
Appendix~\ref{moa_sect} and presented in Fig.~\ref{moa2_fig}.

The eruption of BLG-360 thus followed the principal characteristic of the red transients,
i.e. evolution to progressively lower effective temperature in course of the
eruption and decline as a late M-type (super)giant.

BLG-360 also shares other features observed in the light curves of 
other red transients. The three
peaks in the light curve of BLG-360 with the first one being significantly
fainter than the two subsequent ones (see Fig.~\ref{lc_fig}) are quite
reminiscent of three maxima observed in V838~Mon during its 2002
eruption. One however immediately notes that BLG-360 evolved significantly
slower than V838~Mon. While the peaks in V838~Mon were spanned with roughly
one month, those observed in BLG-360 were separated by seven months. 

\begin{figure}
  \includegraphics[scale=0.4]{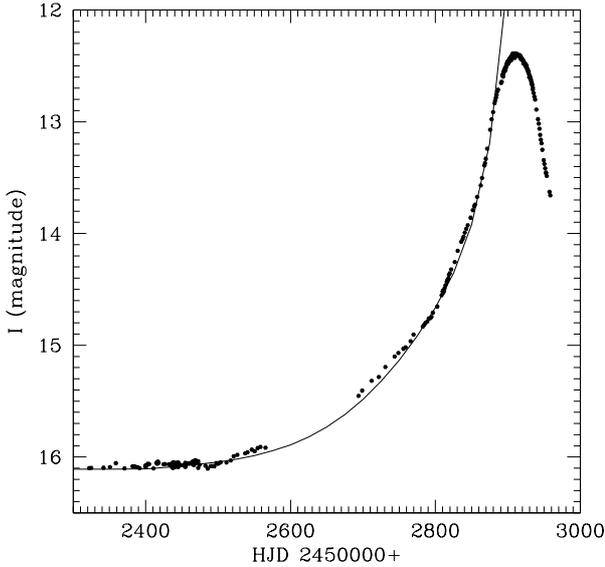}
  \caption{$I$ light curve of OGLE-2002-BLG-360 during its rise to the first maximum.
The line shows an exponential fit using the same formula as in 
\citet[][their Eq.~(2)]{thk11} in their analysis of the rise of V1309~Sco.
}
  \label{rise_fig}
\end{figure}

The shape of the rise of BLG-360 observed in 2002--2003, as displayed in
Fig.~\ref{rise_fig}, was strikingly
similar to that of V1309~Sco observed in 2008 \citep{thk11}. It can be
fitted with the same formula as in \citet[][their Eq.~(2)]{thk11}, although the
time scale used in the formula has to be significantly longer in the case 
of BLG-360. Indeed, the rise in BLG-360 lasted almost 14 months, while that
of V1309~Sco took about 6 months.

Finally, the observational appearance of the remnant of BLG-360 also
argues in favour of the red transient interpretation. As discussed in
Sect.~\ref{lc_sect}, the object disappeared from the optical in 2009.
However, as discussed in detail in Sect.~\ref{sed_rem_sect}, infrared data
show that in 2010 the object remained very bright in the infrared. This is
typical for red transients. The remnants of V838~Mon,
V4332~Sgr and V1309~Sco are all infrared bright or even dominated by 
emission in the infrared \citep{tyl05,kst10,nicholls}.

The eruption of BLG-360 was particularly long and of a low amplitude, when
compared to the other red transients. If the amplitude is defined as a difference
between the maximum brightness during the eruption and the brightness of the
progenitor, then it was of only 5\,mag in the case of BLG-360, compared
to 7--10\,mag observed in V838~Mon, V4332~Sgr, and V1309~Sco
\citep{mun02,martini,thk11}. The
duration of the eruption, as measured at a half amplitude, was of about 26
months in the case of BLG-360.  Those of the other red transients spanned
from 1.5 months in the case of V1309~Sco to 3.5 months for V838~Mon.

\section{Distance and interstellar reddening  \label{dist_sect}}


The photometric data alone, i.e. with no spectroscopic observations of
BLG-360, 
do not allow us to derive any reliable estimate of the distance nor of the
interstellar reddening. The observed position of the object suggests that it
might be a Galactic bulge object. We adopt this as a working hypothesis in
the present paper. We show below that this is not an unrealistic
possibility.

With this hypothesis a rough estimate of the interstellar extinction can be
obtained from the observed reddening of planetary nebulae seen near the
position of BLG-360. These are luminous objects and most of those lying
towards the Galactic bulge are indeed bulge objects. 
Within one square degree around the position of BLG-360, we found
nine planetary nebulae having reliable estimates of the reddening in the
literature \citep{tas92,gcs09}. A mean value and a standard deviation 
derived from these data are $E_{B-V} = 1.03 \pm 0.21$.

The interstellar extinction can also be estimated using the extinction
maps obtained from the OGLE-III survey, as discussed in \citet{nataf}. For
the position of BLG-360, the maps give $A_I$ = 1.515 and $E_{V-I}$ = 1.236,
thus $A_V$ = 2.75. With a standard value of
$R_V = A_V/E_{B-V}$ = 3.1, one gets $E_{B-V}$ = 0.89. Using $R_V \simeq 2.5$,
as found in \citet{nataf}, $E_{B-V}$ increases to 1.10.

Without any chance of getting better estimates we adopt in the present
study that BLG-360 is at a distance of 8.2~kpc 
\citep[Galactic centre distance as determined by][]{nataf}
and is reddened with $E_{B-V} = 1.0$. Following \citet{nataf}, we also adopt 
$R_V = 2.5$.

\section{Spectral energy distribution \label{sed_sect}}  

In archives of several infrared surveys, we have found images and measurements 
of BLG-360. Together with the OGLE measurements this allows us to study the
spectral energy distribution (SED) of the object. 

In all analysed cases,
the object was significantly more luminous in the
infrared than in the optical and shows a clear dust component. Therefore,
to interpret the observations
we have developed a simple model, in which a central star is embedded
in a dusty envelope.

The star is parametrized by its spectral type (or its
effective temperature $T_{\rm star}$) and radius $R_{\rm star}$ 
(or luminosity $L_{\rm star}$).
Its spectrum is interpolated from a set of standard photometric spectra, 
the same as in \citet{tyl05} and \citet{tks11}. 

The dust envelope is spherically symmetric relative to the star and has
a radius $R_{\rm dust}$, and a geometrical thickness $\Delta\,R$.
Dust is uniformly distributed in the envelope
and has an uniform temperature $T_{\rm dust}$. In our modelling we use
$\Delta\,R/R_{\rm dust} = 0.2$ 
(the model results are not particularly sensitive to this parameter). 
Dust grains are assumed to be composed of silicates 
with optical properties taken from the web page of B.\,T.
Draine\footnote{http://www.astro.princeton.edu/$\sim$draine/dust/dust.html}.
Grain sizes follow the standard $a^{-3.5}$ distribution with an upper 
exponential cutoff at 1.0~$\mu$m. The absorption
and scattering coefficients, $Q_{\rm abs}$ and $Q_{\rm sca}$, as well as the
anisotropy factor, $g \equiv \langle \cos \theta \rangle$, are obtained
from integrating the corresponding optical properties over 
the grain-size distribution. 

The stellar radiation is absorbed and scattered in the part of the envelope 
lying in front of the star and having
an optical thickness, $\tau_\lambda^s$. Thus, the stellar SED
is multiplied by $\exp (- \tau_\lambda^s)$. Outward scattering does not
significantly attenuate the stellar radiation, so we defined 
an effective extinction coefficient as
\begin{equation}
 Q_{\rm ext} = Q_{\rm abs} + Q_{\rm sca} (1 - g).
\end{equation}
 The factor $(1 - g)$ roughly accounts for
the isotropic part of the scattering coefficient. 
 The dust envelope is
parametrized by its optical thickness in front of the star
at the effective wavelength of the $V$ band, $\tau_V^s$. Then
$\tau_\lambda^s$ is obtained from normalizing $Q_{\rm ext}$ to $\tau_V^s$ 
at the $V$ band.

In order to calculate the observed emission from the dust envelope, we treat
the envelope as a circular slab or disc of radius 
$R_{\rm dust}+0.5\,\Delta\,R$,
resulting from projection of the
spherical envelope dust distribution on a two-dimensional plane. At a given
position on the disc a local monochromatic intensity, $I_\lambda$, 
is then obtained from
\begin{equation}
 \label{i_eq}
  I_\lambda = B_\lambda(T_{\rm dust})\,[1 - \exp\,(\tau_\lambda)],
\end{equation}
where $B_\lambda$ is the Planck function and $\tau_\lambda$ is the local
optical thickness of the disc. The observed monochromatic flux emitted by the
envelope is then obtained from integrating $I_\lambda$ over the disc
surface and dividing by $d^2$, where $d$ is a distance to the object.

At shorter wavelengths, scattering dominates
absorption. To account for a random walk of photons in scattering
dominated regions, we defined an effective absorption coefficient as 
\begin{eqnarray}
 Q_{\rm abs,eff} & = & \sqrt{Q_{\rm abs} \times Q_{\rm sca}(1- g)}~~~{\rm if} 
 ~~~Q_{\rm sca}(1 - g) > Q_{\rm abs} \nonumber \\
 {\rm or}~~~~~~  & &  \\
 Q_{\rm abs,eff} & = & Q_{\rm abs},~~~{\rm otherwise}.  \nonumber
\end{eqnarray}

The optical thickness of the envelope disc in its centre at the effective
wavelength of the $V$ band, $\tau_V^d$, is related to $\tau_V^s$ through
\begin{equation}
 \tau_V^d = 2\,\tau_V^s \frac{Q_{\rm abs,eff}(V)}{Q_{\rm ext}(V)},
\end{equation}
where $Q...(V)$ stand for appropriate coefficients taken at the effective
wavelength of the $V$ band. The optical thickness of the disc centre at a
given wavelength, $\tau_\lambda^d$, is then calculated from normalizing
$Q_{\rm abs,eff}$ to $\tau_V^d$ at the effective wavelength of the $V$ band.
Finally, the effective thickness at any point of the disc, $\tau_\lambda$,
scales to $\tau_\lambda^d$ as the dust surface density at the given point to
that at the disc centre.

In the above approach there are five free parameters, i.e. $T_{\rm star}$,
$R_{\rm star}$, $R_{\rm dust}$, 
$T_{\rm dust}$, and $\tau_V^s$, which determine the model SED. Their values
can be estimated from fitting the model to the observed photometry. All the models
were calculated assuming radiative equilibrium in the sense that the total
luminosity outgoing from the model must be equal to the intrinsic luminosity of the
central star. 

\subsection{Progenitor \label{sed_prog_sect}}

In the archives of the Two Micron All Sky Survey \citep[2MASS;][]{2mass} we found measurements of 
the progenitor of BLG-360 made on 16~July~1998. The catalogue magnitudes are 
$J = 14.07 \pm 0.15$, $H = 12.65 \pm 0.05$, and $K_s = 11.25 \pm 0.03$.
Images obtained with the Midcourse Space Experiment \citep[MSX;][]{msx} in the A (8.28~$\mu$m) band also show 
a bright object at the position of BLG-360. The observations were made on
31~July~1996. The MSX catalogue gives a flux of 0.15$\pm$0.01\,Jy. 
As discussed in Sect.~\ref{var_sect}, the progenitor of BLG-360 was relatively
stable, at least over a period of a few years before the eruption. 
Therefore we can combine the above
data with the OGLE measurements performed in 2001, i.e. $V = 19.30\pm0.05$
and $I_c = 16.07\pm0.03$ (mean values and standard deviation).

\begin{figure}
  \includegraphics[scale=0.4]{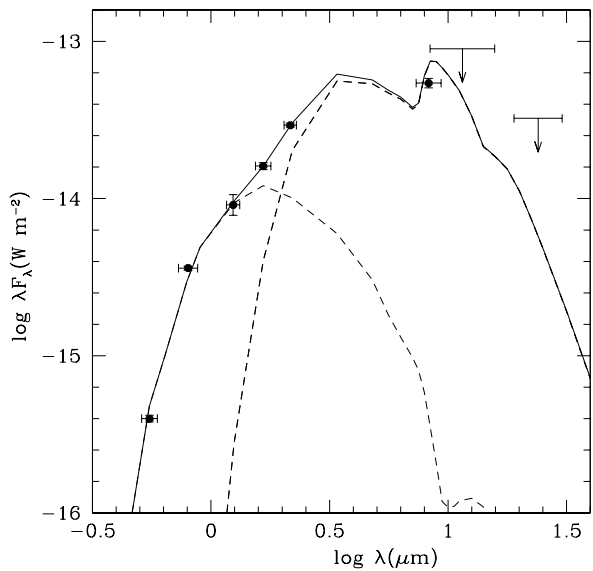}
  \caption{Spectral energy distribution of the progenitor of BLG-360. 
Symbols: observational data from OGLE-III, 2MASS, and MSX, as well as upper limits 
from IRAS (see text for more details). Vertical error bars:
estimated uncertainties of the measurements.
Horizontal error bars:
photometric band widths used in the observations. Full curve: model spectrum
obtained with the values of the model parameters listed in the first row of
Table~\ref{sed_tab}. Dashed curves: individual contributions of the star 
and the dusty envelope.}
  \label{prog_sed_fig}
\end{figure}

Additionally, the position of BLG-360 was observed by the Infrared Astronomical Satellite (IRAS) in 1983. 
The object was not detected but using the IRAS Scan Processing and Integration tool\footnote{http://scanpiops.ipac.caltech.edu:9000/applications/Scanpi/index.html}, 
we were able to put sensitive upper limits on the fluxes of the progenitor 
in the 12~$\mu$m and 25~$\mu$m bands. 
The 3$\times$rms values for the two bands are 0.36 and 0.27\,Jy, respectively. 
Owing to much lower resolution and a presence of strong far-infrared sources close to BLG-360, 
we were not able to estimate useful upper limits for the two other IRAS bands at 60 and 100~$\mu$m.

Fig.~\ref{prog_sed_fig} presents the observational data (symbols) 
and our model of the pre-outburst SED. A giant star of spectral type K\,1 (left dashed curve) 
satisfactorily explains the observed fluxes in $V$, $I$, and $J$. The continuum emission at longer
wavelengths result primarily from the dust envelope (right dashed
curve). Dust has a temperature of 780~K and the envelope attenuates the
stellar radiation with $\tau_V^s = 2.9$. Note that the model
curves are reddened with $E_{B-V} = 1.0$. At a distance of 8.2~kpc the total
luminosity of the object is $\sim$290~L$_\odot$. The dust component
contributes 80\% to the observed luminosity.

\subsection{Eruption  \label{sed_erupt_sect}}

On 24/25 October 2003, BLG-360 was observed in the $K$ band using
the SMARTS 1.3-m telescope at the Cerro Tololo Inter-American Observatory equipped 
with the ANDICAM-IR camera. We performed aperture photometry of the images 
calibrating the magnitude scale using catalogue $K$ magnitudes of field stars 
from 2MASS, DENIS, and VVV surveys. This resulted in $K = 7.25 \pm 0.15$ for 
BLG-360. The same night, BLG-360 was also measured with OGLE with $I = 12.80$. 
For  23/24, 29/30~October and 31~October/1~November we have also OGLE 
measurements of BLG-360 in the $V$ band. Interpolating
from these data, we obtained $V = 16.84$ for 24/25 October. 
The object was then fast declining (see Fig.~\ref{lc_fig}), so we assign 
a conservative uncertainty of 10\% to this photometric point.

\begin{figure}
  \includegraphics[scale=0.4]{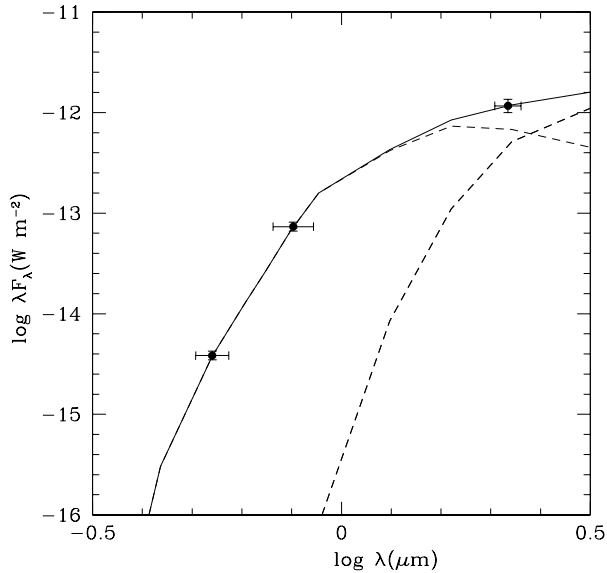}
  \caption{Spectral energy distribution of BLG-360 on 24/25~October~2003. 
Full symbols: measurements from OGLE ($V$ and $I$) and CTIO
($K$) (see text for more details).
Curves: model SED obtained with the parameters listed in the second row of
Table~\ref{sed_tab}.}
  \label{erupt_sed_fig}
\end{figure}

The above data, converted to fluxes, are shown in Fig.~\ref{erupt_sed_fig}.
The three observational points cannot be satisfactorily 
fitted with a single stellar component. A contribution from dust is
necessary to explain the $K$ measurements if
the $V$ and $I$ fluxes are fitted with an M\,3 supergiant.
Having just one point we cannot constrain the dust temperature. But from the
analyses done in Sect.~\ref{sed_prog_sect} and \ref{sed_decl_sect} we found
that it would be reasonable to adopt $T_{\rm dust} \simeq 800$~K.
With this assumption the model envelope has $\tau_V^s = 2.5$ and
contributes 65\% to the observed luminosity of 
$\sim 8.1 \times 10^3$~L$_\odot$.

It is clear that the above model results, as derived from fitting a relatively poor
observational material, should be treated with caution. 

\subsection{Decline \label{sed_decl_sect}}

In the archives of the Spitzer Space Telescope ({\it Spitzer}), we found images 
covering the position of BLG-360 and taken with the InfraRed Array Camera (IRAC) in its all four
channels, i.e. ch1 (3.6~$\mu$m), ch2 (4.5~$\mu$m), ch3 (5.8~$\mu$m), and 
ch4 (8.0~$\mu$m). The observations were obtained on 10--13 May 2007. 
The object was bright in all images and even partly overexposed, 
particularly at longer wavelengths. We measured desaturated fluxes of 
the source using a rectifying procedure of 
T.~Jarrett\footnote{avaible at http://www.ast.uct.ac.za/$\sim$jarrett/irac/tools}. 
This gave us fluxes, $F_\nu$, 
of 4.10, 4.94, 6.30, and 6.02\,Jy for ch1, ch2, ch3, and ch4, respectively.
The object was also observed on 12 October 2006 with the Multiband Imaging Photometer for {\it Spitzer} (MIPS) but only data in the
24~$\mu$m channel turned out to be useful.  Using the same procedure as for IRAC data, we derived
a flux of 1.47\,Jy in this band. All the measurements done on {\it Spitzer} data  
should be accurate to within 10\%. 

The AKARI/IRC All-Sky Survey Point
Source Catalogue \citep[Version 1.0;][]{akari} contains a source at the
position of BLG-360 which was measured in the 18.39\,$\mu$m band with
$F_{\nu}$=2.10$\pm$0.08\,Jy. The source was observed four times between
May 2006 and August 2007 and the given flux is an average value from two
of these observations. The catalogue does not specify
to which particular dates this flux corresponds. Nevertheless we include
the above result in our analysis, although one has to remember that this
measurement could have refered to a somewhat earlier epoch than that of the 
{\it Spitzer} observations. The same
catalogue does not list any source at the position of BLG-360 in the
9\,$\mu$m band of the AKARI's Infrared Camera (IRC), nor we find any
measurements for the four bands of the Far-Infrared Surveyor (65, 90,
140, and 160\,$\mu$m). The future versions of AKARI catalogues, which
will account for variable sources, as well as a release of the original
maps acquired by AKARI should clarify the point.

At the time of the {\it Spitzer} and AKARI observations, BLG-360 was
about two years after its third maximum, initially declining and next 
relatively stable in the $I$ band (see Fig.~\ref{lc_fig}). 
A mean value derived from the OGLE measurements done
between May 2006 and August 2007 is $I = 17.37 \pm 0.42$. No
measurements were done in $V$ in this time period, but from the fact that 
in 2005 the object declined to $V = 20.5-21.0$, we can put a conservative
limit of 21\,mag to the $V$ brightness of BLG-360 in 2006/7.

\begin{figure}
  \includegraphics[scale=0.4]{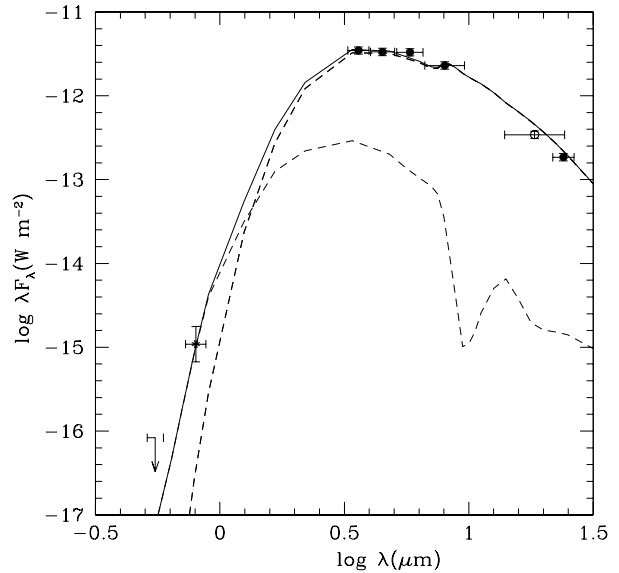}
  \caption{Spectral energy distribution of BLG-360 in May 2006 -- August 2007. 
Symbols: $V$ upper limit and $I$ measurement (asterisk) from OGLE,
{\it Spitzer} IRAC and MIPS measurements (full circles), as well as the
AKARI measurement (open circle). See text for more details.
Curves: model results obtained with the parameters listed in the third row
of Table~\ref{sed_tab}.}
  \label{decl_sed_fig}
\end{figure}

All the above data are presented in Fig.~\ref{decl_sed_fig}. They cover more
than a year of the evolution of the object. In the optical the object was
variable in this time period, i.e. first declining by $\sim$2~mag during
four months and next remaining relatively stable (see Fig.~\ref{lc_fig}). 
However, as can be seen
from Fig.~\ref{decl_sed_fig}, the SED was then
dominated by the infrared emission. It is rather unlikely that the object was
that much variable in the infrared as observed in the optical. Therefore we
believe that the global parameters we derive from the model fitting to the
observations are typical for the late decline of BLG-360.

As mentioned above,
the SED of BLG-360 in 2006/7 was dominated by the
infrared component. The data from {\it Spitzer} and AKARI can be 
satisfactorily reproduced 
by a dust component with $T_{\rm dust} \simeq 850$~K and a luminosity of 
$\sim 1.2 \times 10^4$~L$_\odot$. This component cannot however explain
the observed brightness in $I$. We therefore assume that the
central object was still seen in the optical although it must have been heavily
obscured by circumstellar dust responsible for the infrared brightness.
We therefore add, in our modelling, an M\,4 supergiant (this spectral type is
somewhat arbitrary) of $L = 1.25 \times 10^4$~L$_\odot$.
The central star is attenuated with $\tau_V^s = 8.5$. 
The curves in Fig.~\ref{decl_sed_fig} show the results of
the above modelling. For an observer, the star contributes only 7\% to the
total observed luminosity of the object.

\subsection{Remnant \label{sed_rem_sect}}

In 2010, i.e. about three years after BLG-360 disappeared from the optical,
the object was observed by the Wide-field Infrared Survey Explorer \citep[WISE;][]{wise} 
and within the Vista Variables in the Via Lactea (VVV) ESO Public Survey \citep{vvv}. 

WISE observations of BLG-360 were made on 19--21~March~2010.
Images were taken in all four bands, i.e. W1 (3.35~$\mu$m), W2 (4.6~$\mu$m), W3 (11.6~$\mu$m),
and W4 (22.1~$\mu$m), and they all show a bright object at the position of 
BLG-360. From the WISE catalogue, we calculated colour-corrected fluxes of 
0.85, 4.39, 5.32, and 3.50\,Jy
for the bands W1, W2, W3, and W4, respectively. These values are uncertain
to $\sim$5\,\%.

The Data Release 1 of the  VVV survey also provided us with images 
covering the position of BLG-360. 
The thus-far available observations were obtained in August--September 2010. 
The images taken with
the $Z$ (0.89~$\mu$m) and $Y$ (1.02~$\mu$m) filters do not show any
measurable object at the position of BLG-360. The object is however clearly
seen in the $H$ (1.65~$\mu$m) and $K_S$ (2.15~$\mu$m) bands. 
Using catalogue products associated with the survey, 
we estimated the fluxes of 3.1 and\,29.5 mJy, with their corresponding
uncertainties of 10\,\% and 3\,\% in the latter two bands,
respectively.

\begin{figure}
  \includegraphics[scale=0.4]{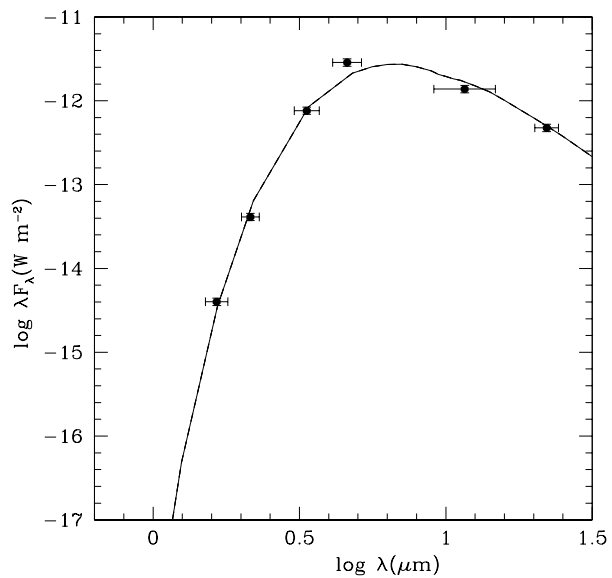}
  \caption{Spectral energy distribution of the remnant of BLG-360 in 2010. 
Full symbols: measurements from the VVV ($H$ and $K_S$) and WISE surveys in 
March--September 2010.
Full curve:
an optically thick dust model with $T_{\rm dust} = 550$~K.}
  \label{rem_sed_fig}
\end{figure}

All the above post-outburst measurements are shown in Fig.~\ref{rem_sed_fig}.
The full curve in the figure presents a model dust spectrum fitted to
the WISE and VVV data. The dust has a temperature of 550~K
and is assumed to be optically thick,
i.e. $\tau_V^s = 100$. The need for high optical thickness is motivated by 
the fact that neither optical (OGLE-IV) nor 
near-infrared ($Z$ and $Y$ in VVV) observations revealed any object at the
position of BLG-360. To satisfy the latter we found that $\tau_V^s \ga 20$.
The remnant of BLG-360, if modelled as in Fig.~\ref{rem_sed_fig},
has a luminosity of $8.2 \times 10^3$~L$_\odot$.

\subsection{Summary  of the SED analysis \label{sed_sum_sect}}

Table \ref{sed_tab} summarizes the parameters of the stellar and dust
components that we derived from the analysis of the observed SED made in the
previous subsections. First two columns give the dates and the sources of the
observational data used to construct the SED. Next four columns present the
stellar parameters, i.e. spectral type, effective temperature, as well as 
radius and luminosity in solar values. The subsequent column gives the
optical thickness in the $V$ band of the dust envelope 
in the line of sight of the star. Last
four columns present parameters of the dust envelope responsible for the
infrared emission, i.e. dust temperature, luminosity in terms of the stellar
value, as well as radius, and dust
mass in terms of the solar values. 
A dust grain specific density of 3.3~g\,cm$^{-3}$ was adopted when
evaluating the dust mass.

\begin{table*}
\begin{minipage}[t]{\hsize}
\caption{Basic parametres of BLG-360 derived from the SED analysis}
\label{sed_tab}
\centering 
\renewcommand{\footnoterule}{}  
\begin{tabular}{@{}c@{}c@{}ccrrrccrr}
\hline
\hline
 date & data source & sp.type & $T_{\rm star}$(K) & \multicolumn{1}{c}{$R_{\rm star}/{\rm R}_\odot$} &
 \multicolumn{1}{c}{$L_{\rm star}/{\rm L}_\odot$} & $\tau_V^s$ & $T_{\rm dust}$(K) & 
 $L_{\rm dust}/L_{\rm star}$ & \multicolumn{1}{c}{$R_{\rm dust}/{\rm R}_\odot$} &
 \multicolumn{1}{c}{$M_{\rm dust}/{\rm M}_\odot$} \\
\hline
 1996--2001 & OGLE,\,2MASS,\,MSX & K\,1 & 4\,350 & 30 & 290 & 2.9 & 780 
  & 0.80 & 1\,070 & $2.9 \times 10^{-8}$ \\
 Oct 2003 & OGLE,\,CTIO & M\,3 & 3\,600 & 230 & 8\,100 & 2.5 &
   800\footnote{assumed} & 0.65 & 5\,000 & $5.5 \times 10^{-7}$ \\
 2006--2007 & {\it Spitzer},\,AKARI,\,OGLE & M\,4 & 3\,500 & 300 & 12\,500 & 8.5
  & 850 & 0.93 & 5\,200 & $1.8 \times 10^{-6}$ \\
 Mar--Sept 2010 & WISE,\,VVV & M\,6--7 & $3\,200^a$ & 300 & 8\,200 &
  $\ga 20$ & 550 & 1.00 & 10\,000 & $\ga 1.4 \times 10^{-5}$ \\
\hline
\end{tabular} 
\end{minipage}
\end{table*}

Note that in the SED model for October~2003 (second row in
Table~\ref{sed_tab}) the dust temperature 
is an assumed value (see Sect.~\ref{sed_erupt_sect}). Consequently the
dust envelope parameters derived in this case should be treated with
caution. Similarly, the stellar effective temperature (spectral type) 
in the last row of Table~\ref{sed_tab} is also an assumed value.
In 2010, the central star was not visible and the value of
$T_{\rm star}$ given in the table follows the observed spectral types 
of other red transient remnants (V838~Mon, V4332~Sgr, V1309~Sco).
Finally, note that the whole SED analysis was made assuming a distance of
8.2~kpc and an interstellar extinction of $E_{B-V} = 1.0$ ($R_V =2.5$) (see
Sect.~\ref{dist_sect}). Thus the derived radii and luminosities would
scale with distance and distance squared, respectively. The star
temperature (spectral type) depends on the extinction but not very sensitively. 
The dust temperature is practically insensitive to the assumed interstellar 
extinction.

\section{Variability of the progenitor \label{var_sect}}

Encouraged by the case of V1309~Sco, whose progenitor showed pronounced variability,
which allowed \citet{thk11} to study the nature of this object and identify the
mechanism of the red transient eruptions, we looked for 
variability of the progenitor of BLG-360, as well. The
OGLE-III monitoring was not sufficiently long before the eruption to provide
us with a conclusive observational material for this purpose. Less than two
years of the OGLE monitoring of the progenitor (see Fig.~\ref{lc_fig}) show
however that the object was not particularly variable just before 
the beginning of the eruption in 2002 (but see below).
Fortunately BLG-360 was also monitored in the MACHO and MOA projects, which
provided us with observational data for epochs earlier than the beginning of
the OGLE-III observations. 

The results from the MACHO project are presented
and analysed in Appendix~\ref{macho_sect}. These data cover the period of
1995--99. Although the mean brightness of the object remained stable over
the time span of the observations, the object seemed to show 
variability of small amplitude, which could be interpreted with a periodicity
of $\sim$240~days.
Various reasons can be invoked to explain the variability, i.e. binarity,
pulsation, or rotation of the progenitor. Quasi-periodic variations in the
observed dust-forming mass loss (see Sect.~\ref{disc_prog_sect}) could also
result in variable screening of the central star. Low quality of the observational
data does not allow us to answer if this variability was really present and
what could have been its nature.

\begin{figure}
  \includegraphics[scale=0.4]{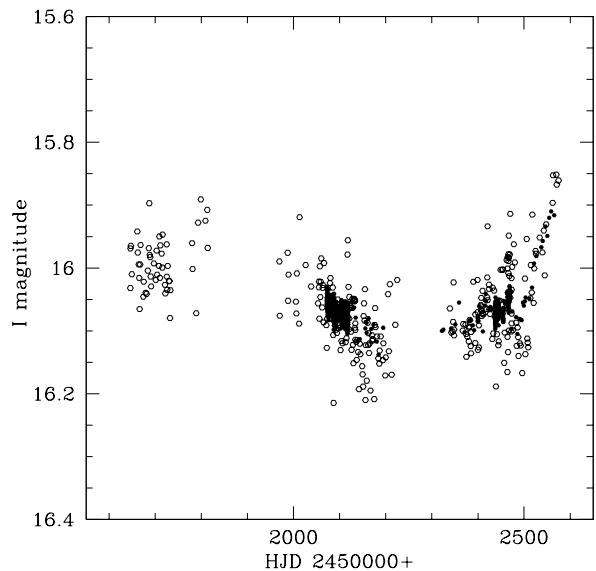}
  \caption{Evolution of the $I$ brightness of BLG-360 in 2000--2002 as 
obtained from the MOA archives (open circles) and the OGLE-III data (full
points). Note that the reference flux in the MOA data was modified, as
explained in Appendix~\ref{moa_sect}.}
  \label{prog_var_fig}
\end{figure}

The photometric data from the MOA project are presented and analysed in
Appendix~\ref{moa_sect}. The data cover the period of 2000--2005, thus they
allow us to study the object in the time period between the end of the MACHO
observations and the beginning of the OGLE-III. Unfortunately, there is a
systematic problem with the MOA data, particularly important in the
progenitor phase, as discussed in 
Appendix~\ref{moa_sect}. We postulate that the reference flux in the MOA
data was significantly overestimated. If corrected to match the OGLE light curve
(see Appendix~\ref{moa_sect}) they can be used together with 
the OGLE results to analyse the progenitor light curve in 2000--2002, as
presented in Fig.~\ref{prog_var_fig}. Although the scatter of the
observational points is significant, particularly in the case
of the MOA data (open circles in the figure), Fig.~\ref{prog_var_fig} shows
that a year before the beginning of the rise, i.e. in 2001, the object
dropped by $\sim$0.1~mag compared to the mean brightness in 2000. This
resembles the behaviour of V1309~Sco before its eruption 
\citep[see Fig.~1 in][]{thk11}, although in that case the drop was more
significant ($\sim$1~mag).

In summary, the progenitor of BLG-360 was not showing any spectacular
variability of the sort observed in the case of V1309~Sco. The 240~days
periodicity, suggested by the MACHO data, is not certain, mostly due to low 
precision of the data. The deepening before the start of the
eruption, reminiscent of a similar phenomenon observed in V1309~Sco, is
a weak effect and in fact not entirely evident due to uncertainties in the
MOA data.

In the digitalized plates from the ESO Red Atlas obtained with the 1-m 
Schmidt Telescope at la Silla Observatory \citep{eso-red}, 
we have found an image taken in August 1980 in the $R_F$ filter,
which shows a stellar object at the position of BLG-360. Our aperture
photometry of the object and field stars, which have measurements in
the Guide Star Catalogue 2 (GSC-2), gave $R_F = 17.4 \pm 0.6$ for the BLG-360
progenitor. It should however be noted that the GSC-2 standards 
are brighter than $\sim$16.3 mag, so our measurement of BLG-360 assumes an
extrapolated calibration, hence the big error. An interpolation between the
$V$ and $I$ measurements from OGLE (see Sect.~\ref{sed_prog_sect}) 
gives $\sim$17.9 mag for the effective wavelength of the $R_F$ filter
($\sim$650~nm). This can be compared to the red magnitude of $\sim$18.1
measured in MACHO (see Appendix~\ref{macho_sect}). 
Given the low angular resolution of the 1980
image, which implies that faint nearby field stars are also included in our
measurement from this image, we can conclude that the progenitor of BLG-360
in 1980 was not significantly different in brightness from that observed
just before the eruption.

We also found scanned plates from the Palomar Observatory Sky Survey 
(POSS, red filter) and from the UK Schmidt Telescope (SRC plates,  
waveband $B_J$) which cover the position of the OGLE 
transient\footnote{Images were retrieved using the service available 
at http://www.cai-mama.obspm.fr/mama/}. 
These observations were obtained in April 1958 and June 1974, respectively. 
They are of too poor quality to perform any useful quantitative photometry 
of BLG-360 in its crowded field, but by comparing them to the images of more 
recent observations smoothed to a similar angular resolution, we can exclude 
any significant variability of the object at these dates, i.e. the object did 
not experience an eruption of the type observed by OGLE 2002--2006 in those earlier epochs.

\section{Discussion \label{disc_sect}}

\subsection{Progenitor \label{disc_prog_sect}}

The SED analysis made in Sect.~\ref{sed_prog_sect} shows that the progenitor
of BLG-360 looked as a K-type giant, $\sim$5~times overluminous as compared to
the standard calibration of luminosity vs. spectral-type \citep{sk82,allen},
surrounded by a dusty envelope, which reprocessed $\sim$80\% of the stellar
luminosity. The dust optical thickness along the line of sight of
the star was $\sim$2.9 in the $V$ photometric band. With the optical
properties of the model dust grains used in our SED modelling we obtain that
the above value of optical thickness corresponds to a dust column density of 
$\Sigma_{\rm dust} \simeq 4.1 \times 10^{-4}$~g\,cm$^{-2}$. 

The dust was hot ($\sim$780~K), which suggests that it was
formed in a stellar wind of the progenitor of BLG-360.
Assuming a steady-state
spherically-symmetric wind, its dust mass-loss rate can be estimated from
\begin{equation}
 \label{m_dot_eq}
  \dot{M}_{\rm dust} = 4\,\pi\,r_0\,\varv_{\rm wind}\,\Sigma_{\rm dust},
\end{equation}
where $r_0$ is a radius of the base of the dusty wind and $\varv_{\rm wind}$ is
a wind velocity. Taking $r_0 = 1070$~R$_\odot$ 
(see Table~\ref{sed_tab}) and $\varv_{\rm wind} = 100$~km\,s$^{-1}$ 
(which is of the order of the escape velocity from a $\sim$1~M$_\odot$ star with 
$R_{\rm star} = 30$~R$_\odot$ -- see first row in Table~\ref{sed_tab}) one
obtains $\dot{M}_{\rm dust} = 6 \times 10^{-8}$~M$_\odot$\,yr$^{-1}$.
With a standard dust-to-gas mass ratio of 0.01 we can conclude that
the mass-loss rate from the progenitor of BLG-360 was of the order of
$10^{-5}$~M$_\odot$\,yr$^{-1}$.

This mass-loss rate is orders of magnitude too high as for a typical
red giant. Therefore it is rather impossible that the progenitor of BLG-360
was a more or less normal red giant, no matter whether being a single star or
a primary (brighter) component of a binary system. One could suggest 
that the object was
an asymptotic giant branch star. However, in this case the object
would be at least an order of magnitude more luminous than the value
given in the first row of Table~\ref{sed_tab}. This would imply a distance
at least three times larger than that adopted in Sect.~\ref{dist_sect}. Thus
the object would have to be situated in outskirts of the Galaxy and seen 
through the Galactic bulge, which seems unlikely.

The strong similarity of the rising phase of the BLG-360 eruption, shown
in Fig.~\ref{rise_fig}, to the same phase of V1309~Sco \citep{thk11} might
suggest a similar nature of the progenitor in both cases. However, contrary
to V1309~Sco, the progenitor of BLG-360 did not show any photometric sign 
of being a contact binary. One can postulate that we observed a more or less
pole-on binary. This, however, rises a problem with explaining the
strong infrared excess of the progenitor. Contact binaries do not show that
huge mass-loss rates as the value derived above. Moreover, the luminosity of
$\sim$290~L$_\odot$, as derived in Sect.~\ref{sed_prog_sect}, is also far
too high for a typical contact binary. The latter could be simply solved by
saying that the distance of BLG-360 is much shorter than the value adopted
in Sect.~\ref{dist_sect}. A factor of ten shorter distance would bring the
luminosity to an acceptable value. This would not, however, solve the problem
of the mass-loss rate. As can be seen from  Eq.~(\ref{m_dot_eq}), 
the mass-loss rate would decrease only by a factor of 10, 
still remaining on an unacceptably high level. One can invoke that the
infrared excess is not a sign of a strong, more or less spherical, 
mass loss but is produced by a
disc-like or torus-like dusty envelope formed by mass outflow in the
equatorial plane of the binary. This is what is expected in the case of a
contact binary approaching its merger \citep[as discussed in][]{thk11}.
However in this case the central object would be directly seen (no
significant extinction of the pole-on binary from the envelope) 
and it would be difficult to
understand, why the envelope was $\sim$4 times brighter than the central
binary (see Sect.~\ref{sed_prog_sect}).

The above problems related to the strong infrared excess and the lack of any
significant photometric variability over a time period of at least 7--8 years 
before the eruption can be solved if we postulate that the
progenitor of BLG-360 was a common-envelope binary, namely a binary system
that had entered the common-envelope phase a certain time (at least $\sim$10
years) ago. The scenario could have been as follows \citep[for more details
see, e.g.][]{ivan}. The more massive companion of
a binary filled its Roche lobe when evolving from the main sequence toward the 
red giant branch. This resulted in an initially fast mass transfer and formation
of a common envelope. After a relatively violent initial phase of the common
envelope (plunge-in), the object entered a rather gentle phase (spiral-in) 
when the components were slowly spiralling in inside the envelope. 
This would be the phase, or rather its
end, that we observed in 1995--2001. The binary, deeply embedded in the
envelope, was not directly seen, which explains the lack of any significant
variability of the object. A slow dissipation of the orbital energy
and the related transfer of the orbital momentum to the envelope would have 
resulted in the intense mass outflow. 


\subsection{Eruption  \label{disc_erupt_sect}}

The spiral-in process apparently accelerated in 2002, which resulted in an
exponential rise of the object brightness lasting $\sim$14 months (see
Figs.~\ref{lc_fig} and \ref{rise_fig}). This most likely led to a final merger
of the binary components, or rather of their cores, in 2004.

The light curve of BLG-360 (see Fig.~\ref{lc_fig}) shows three pronounced
peaks during the eruption. One can even claim that the phenomenon
also repeated in the decline: a sort of slow-down in the decline  in 2005 
and a clear halt and even rise in brightness in 2007. A similar phenomenon
of multiple peaks
was also observed in the case of the 2002 eruption of V838~Mon \citep[see
e.g.][]{mun02,crause03}. The origin of the phenomenon is not
clear. In the case of V838~Mon, \citet{tylsok06} proposed that the peaks
could have been reminiscence of a violent merger of a binary with a highly
eccentric orbit. This is not likely to be the case of BLG-360, given
our analysis of the pregenitor in Sect.~\ref{disc_prog_sect}. The slow
and smooth rise of BLG-360 to its maximum, lasting 14 months, 
shows that the merger process started
well before the first maximum and that from its very beginning the binary
was deep in the common envelope. Perhaps the multiple peaks result from a
certain sort of instability appearing at the very merger and resulting 
in a quasi periodic oscillations of the merger envelope.

BLG-360 increased its $I$ brightness by a factor of 100 when its maximum is
compared to the brightness of the progenitor. If the same factor applies to
the luminosity, then the object would reach $\sim 3 \times 10^4$~L$_\odot$.
The general tendency of the object to become progressively redder in the
$V-I$ colour (see Fig.~\ref{VI_fig}) however implies a progressively
higher bolometric correction -- thus an increase of the luminosity much greater
than that of the $I$ brightness. Unfortunately, the lack of infrared
measurements during the eruption does not allow us to study
this important point in detail. Our attempt to investigate the SED in
October~2003 (see Sect.~\ref{sed_erupt_sect}) gave a value of $\sim 8
\times 10^3$~L$_\odot$. This is however an uncertain value as based only on
measurements in the $V$, $I$, and $K$ bands, while a significant part of
the expected flux in the infrared escaped measurements. The object was then
in a significant decline from the first maximum and much below its maximum
in 2004. A similar analysis made on the data obtained in 2007 (see
Sect.~\ref{sed_decl_sect}) gave a more reliable value of 
$\sim 1.3 \times 10^4$~L$_\odot$. But the object was then in a deep decline,
i.e. $\sim$6\,mag in $I$ below its maximum brightness. If the same factor
is applied to the luminosity then we get $\sim 3 \times 10^6$~L$_\odot$ in
maximum. This value is however rather
an upper limit to the maximum luminosity since the observed decline in $I$ was
probably not only due to the decline in luminosity but also due to the reddening of
the object and dust formation in the outflowing matter. It seems reasonable
to conclude that during the eruption BLG-360 increased in luminosity 
at least by a factor of 100, but a factor of 1000 cannot be excluded. Its
maximum luminosity was probably of a few $\times 10^4$~L$_\odot$.

\subsection{Remnant  \label{disc_rem_sect}}

The luminosity of the BLG-360 remnant in 2010 is relatively well determined (see
Sect.~\ref{sed_rem_sect}), more or less at the same level of accuracy as
that in 2007. During three years the object declined 35\% in luminosity 
(see Table~\ref{sed_tab}).
The remnant is now completely embedded in dust. The dimensions of the dusty
envelope are of $\sim 1 \times 10^4$~R$_\odot$ (see Table~\ref{sed_tab}),
i.e. some 50~AU. It expanded by a factor of 1.9 between 2007 and 2010,
which corresponds to an expansion velocity of $\sim$35~km\,s$^{-1}$. This
measures the expansion of a sort of effective photosphere of the dusty
envelope, so it need not to be the expansion velocity of the outflowing
matter. Nevertheless, the above value is close to the escape velocity from a
solar-mass star inflated to $\sim$300~R$_\odot$ (see
Table~\ref{sed_tab}).

As can be seen from Table~\ref{sed_tab} (last column), the mass of dust
seen in the infrared was increasing with time. This can be interpreted as
evidence of an increasing mass of the outflow during the eruption. The value
obtained from the observations in 2010 can be considered as a measure of
the total mass lost during the eruption. The dust envelope was optically
thick in 2010 so we can derive only a lower limit to the mass.
If we adopt a standard value of 0.01 of the
dust-to-gas mass ratio, then we can conclude that BLG-360 lost at least
$10^{-3}$~M$_\odot$ during its eruption.

\subsection{CK Vul: an analogue to BLG-360?  \label{ckvul_sect}}

CK Vul, also known as Nova Vul 1670, is usually classified as 
a slow classical nova \citep[e.g.][]{shara85} or a final He-shell flash
post-AGB object \citep[e.g.][]{evans02}. 
\citet{kato} however proposed, following the merger scenario
for V838~Mon of \citet{soktyl03}, that the eruption of CK~Vul might have
also resulted from a stellar merger. The main argument based on similarities
in the light curve of CK~Vul and V838~Mon.
The idea was not however overwhelmingly accepted, mostly due to the time
scale of the eruption, which in the case of CK~Vul was ten times longer than
in V838~Mon.

\begin{figure}
  \includegraphics[scale=0.4]{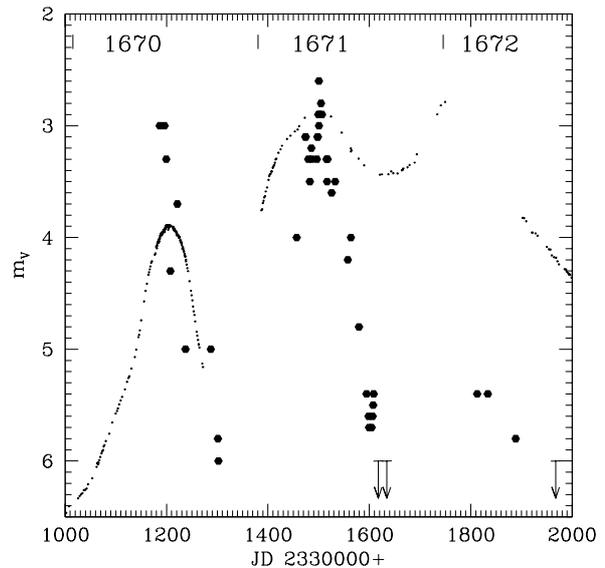}
  \caption{Visual light curve of CK~Vul (Nova Vul 1670) compared to the $I$
curve of BLG-360. Full symbols: eye estimates of CK Vul in 1670--1672 
\citep[the data are from][arrows: upper limits]{shara85}. 
Small dots: $I$ measurements of BLG-360 in
2003--2005. Notes: the BLG-360 light curve is shifted in the magnitude
scale to match the brightness of CK~Vul, the time scale of BLG-360 is
increased by a factor of 1.4.
}
  \label{CKVul_fig}
\end{figure}

Figure \ref{CKVul_fig} compares the light curve of CK~Vul to that of
BLG-360. The results of eye estimates of the CK~Vul brightness was taken
from \citet{shara85}. The $I$ measurements of BLG-360 were shifted in the
magnitude scale to match, more or less, the observed magnitudes of CK~Vul.
Also the time scale of BLG-360 was multiplied by 1.4, to get a 
spacing between the maxima similar to that in CK~Vul.

As can be seen from Fig.~\ref{CKVul_fig}, in the case of CK~Vul as well as in 
BLG-360, three distinctive peaks are clearly present
in the light curve. The time span between the peaks is also similar in both cases, 
i.e. ten months in CK~Vul versus seven months in BLG-360. Thus the light
curve of BLG-360 is much closer to that of CK~Vul than it was in the case of
V838~Mon. 

Of course there are clear differences between the light curves of both
objects. The third maximum is fainter and the minimum between the second and
third peaks is deeper in CK~Vul than in BLG-360. Part of these differences
can be atributed to the differences in colour sensitivity between the eye and
the $I$ filter, especially if the object becomes very red.

Futher similarities between both objects can be found in observational appearances 
of the remnants. Both, in the case of CK~Vul and BLG-360, no remnant is visible in
the optical. Although an emission nebula was found arround the position of
CK~Vul, no optical stellar-like object has so far been identified
\citep[e.g.][]{shara85}. However, similarily as in the case of BLG-360, 
a bright infrared source is clearly seen at the position of CK~Vul
\citep{evans02}.

We thus conclude that the case of BLG-360 provided evidences favouring the
hypothesis that the eruption of CK~Vul observed in 1670-72 was of a red-transient
type.

\subsection{Similarities to NGC\,300~OT2008  \label{ngc300_sect}}

The progenitor of BLG-360 showed a strong infrared excess, already at short
wavelengths ($JHK$), indicating that it
was an evolved object with an intense mass-loss. In this respect the object
was different from the other Galactic red transients.
The photometric data on the progenitor of V838~Mon
do not show any infrared excess up to the $K$ band \citep{tss05}. 
In the case of V4332~Sgr the situation is less clear. No near infrared
measurements are available for the progenitor and the object
was too faint to be detected in the IRAS survey. The
progenitor of V1309~Sco was not a bright infrared source \citep{nicholls}.

From the point of view of the infrared behaviour, BLG-360
shows clear similarities to the extragalactic optical transient NGC\,300~OT2008. 
The progenitor of the latter object was identified as a luminous infrared
source \citep{prieto08,berger09}.
Similarly as BLG-360, the object also showed a pronounced infrared excess during 
the eruption \citep{prieto09} and became completely embedded in dust after 
the outburst \citep{prieto10}. It is thus tempting to suggest that the progenitor of
NGC\,300~OT2008 could also have been a common-envelope binary and that its
final merger resulted in the observed outburst.

NGC\,300~OT2008 was significantly brighter than BLG-360. At maximum the
object reached $5 \times 10^6$~L$_\odot$, compared to a few
$\times 10^4$~L$_\odot$ if BLG-360 is a Galactic bulge object. A similar
factor also differs the brightnesses of the progenitors. The time scale of
the eruption of NGC\,300~OT2008 was however a factor of 10 shorter than that
of BLG-360. The above differences can be accounted for by differences in
masses. The progenitor of NGC\,300~OT2008 was a young massive
(binary) object \citep{bond09,berger09}, while that of BLG-360 was probably an old 
low-mass Galactic bulge object. A denser common envelope in the former case
could explain the more luminous but shorter eruption.

\section{Summary}
OGLE-2002-BLG-360, originally discovered as a gravitational microlensing candidate,
appears to be an overlooked red transient. This interpretation is primarily
supported by the observed evolution of the $V-I$ colour, which shows that
the object became progressively cooler in course of the eruption and the
decline. In the end, the object disappeared from the optical, remaining
bright in the infrared. Clearly the remnant is now completely imbedded in
dust formed in matter lost by the object during its eruption.

The eruption of BLG-360 is, in some aspects, different from those of the
other Galactic red transients, i.e. V838~Mon, V4332~Sgr, and V1309~Sco. First, it
was less violent, in the sense that the amplitude of the outburst was only
5~mag, compared to 7--10 mag in the other objects. Second, it lasted much
longer, i.e. about 3~years compared to months in the other cases. The light
curve of BLG-360 and the duration of its eruption bears however strong
similarities to CK~Vul erupted in 1670-2, strengthenning the idea that the
latter was also of the red-transient type. We point out similarities in the
infrared evolution of BLG-360 and the extragalatic optical transient
NGC\,360\,OT2008. We thus speculate that the latter event could be of a
similar nature as the eruption of BLG-360.

We interpred the observations of BLG-360 in terms of a merger of an
evolved low-mass binary. The lack of any significant variablilty of the
progenitor over decades before the eruption, as well as the strong infrared
excess, implying an intense mass-loss, can be understood if we postulate
that the progenitor was a common-envelope binary being in 
the so-called spiral-in phase. This phase ended in 2002 when the cores of   
the binary components entered a violant merger phase, which resulted in the 
observed eruption. Copious dust-condensation in matter outflowing from the  
merging binary finally hiddened the object, so it became invisible in the   
optical. After a certain time, when dust partly disperses in space, we will 
perhaps have a chance to see the central star radiation scattered on denser 
dusty regions. The radiation would also populate low-lying excitation levels of
atoms and molecules in these regions. Thus it is possible that, in a certain
future, the object reappears in the optical with a
spectrum resembling that of the V4332~Sgr remnant \citep{kst10}.

\begin{acknowledgements}
The OGLE project has received funding from the European Research Council under 
the European Community's Seventh Framework Programme (FP7/2007-2013) / ERC 
grant agreement no. 246678. The research reported in this paper has partly been 
supported by a grant no. N\,N203\,403939 financed by the Polish Ministry of Sciences 
and Higher Education. We thank Marcin Hajduk for participating in
photometric measurements of the $K$-band images obtained at CTIO on 
24/25 October 2003 (Sect.~\ref{sed_erupt_sect}) and the
digitailsed plate from the ESO Red Atlas obtained in August 1980
(Sect.~\ref{var_sect}). Thanks also to Richard Pogge for finding out the
CTIO $K$-band observations of BLG-360 obtained in October~2003.
This publication makes use of several data products: 
from the Two Micron All Sky Survey, which is a joint project of the University of Massachusetts and the Infrared Processing and Analysis Center/California Institute of Technology, funded by the National Aeronautics and Space Administration and the National Science Foundation; 
from the Midcourse Space Experiment (the MSX data funded by the Ballistic Missile Defense Organization with additional support from NASA Office of Space Science);
from the NASA/ IPAC Infrared Science Archive, which is operated by the Jet Propulsion Laboratory, California Institute of Technology, under contract with the National Aeronautics and Space Administration (NASA); 
from the Spitzer Space Telescope, which is operated by the Jet Propulsion Laboratory, California Institute of Technology under a contract with NASA; 
from the Wide-field Infrared Survey Explorer, which is a joint project of the University of California, Los Angeles, and the Jet Propulsion Laboratory/California Institute of Technology, funded by NASA. 
from the Cambridge Astronomical Survey Unit;
and data collected at the Paranal Observatory under ESO programme ID 179.B-2002. 
We acknowledge all these institutions.
\end{acknowledgements}

\begin{appendix}
\section{OGLE-2002-BLG-360 in the archives of the MOA project}
\label{moa_sect}

\begin{figure}
  \includegraphics[scale=0.4]{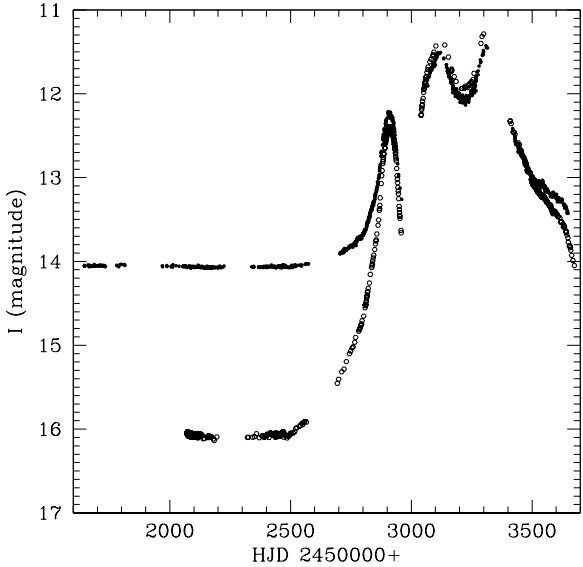}
  \caption{$I$ magnitudes of BLG-360 as obtained from the MOA archives (full
points) compared to the data from OGLE-III (open circles).}
  \label{moa1_fig}
\end{figure}

OGLE-2002-BLG-360 was monitored in the MOA (Microlensing Observations in
Astrophysics) project during the time period of 2000--2005. 
In the archives of MOA the object is named as
MOA-2003-BLG-8\footnote{https://it019909.massey.ac.nz/moa/alert2000/moa-2003-blg-8.html}.
The observations were made with the wide red MOA filter and the data are available
as counts relative to a reference flux. Using Eq.~(7~for~CCD\,1) in \citet{bond_moa}
the data can be transformed into magnitudes in the $I_c$ passband. The
results are presented in Fig.~\ref{moa1_fig} as full points. For comparison
the $I$ magnitides from OGLE-III are also shown in the figure as open
circles.

As can be seen from Fig.~\ref{moa1_fig}, the agreement of the MOA results
near the maxima of the eruption is reasonable. However before the eruption
and in the rising phase the data disagree. The MOA data gives a brightness
of the progenitor at a level of $I \simeq 14.05$, while in the OGLE data
the object is $\sim$2 magnitudes fainter. In the MOA data the initial rise is much
slower and the amplitude of the eruption is significantly smaller than in
OGLE. 

\begin{figure}
  \includegraphics[scale=0.4]{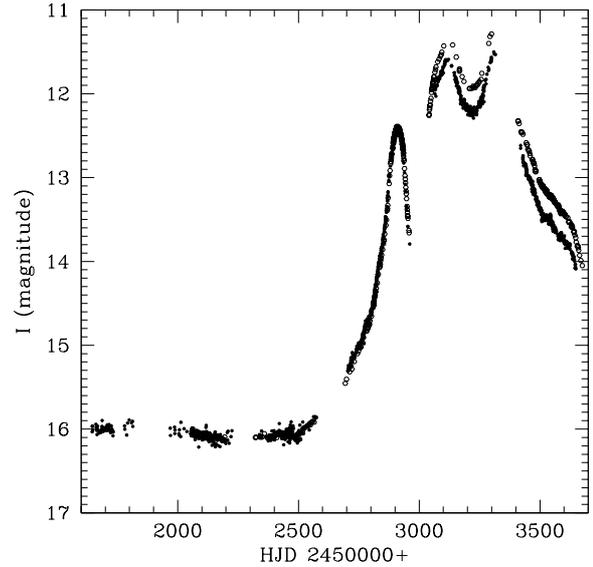}
  \caption{The same as in Fig.~\ref{moa1_fig} but with a modified reference
flux in the MOA data, as explained in the text.}
  \label{moa2_fig}
\end{figure}

We analysed this disagreement in different ways and we conclude
that the reason probably lies in a wrong value of the reference flux in the MOA
archive data. We adjusted the reference flux as to get a good agreement
between MOA and OGLE in the progenitor phase (see Fig.~\ref{prog_var_fig}). 
The results are presented in Fig.~\ref{moa2_fig}. 
The general agreement of the light curves is now very good. 
Small systematic differences can be understood in terms of differences 
in the filters and the evolution of the colour of BLG-360. 
The red filter of MOA is more sensitive
to shorter wavelengths than the $I$ filter of OGLE. As can be seen from
Fig.~\ref{VI_fig}, in the rising phase in 2003, the $V-I$ colour was slightly
bluer than that of the progenitor. This is consistent with the MOA magnitude
beeing then slightly but systematicaly brighter than the OGLE value. On the other
hand, the strong and systematic reddening of the object in the $V-I$ colour 
during the eruption explains while during the 2004 maxima and the 2005
decline the object becomes systematically fainter in MOA  than in OGLE.

The point is, however, that in order to get the agreement between MOA and
OGLE, as shown in Fig.~\ref{moa2_fig}, we had to decrease the reference flux
in the MOA data by 85\%. It is not clear, how to explain this factor. The
angular resolution of the MOA images was typically 2.5\,arcsec
\citep{bond_moa}, which is significantly worse than that of OGLE-III 
(0.7--1.0~arcsec). Hence in a dense field measurements with MOA could
have included more light from field stars compared to OGLE. Indeed the
$I$ OGLE-III images show that there are stars within 1--2\,arcsec from 
BLG-360 but they are faint, significantly fainter that the progenitor
of BLG-360. Thus they cannot explain why the progenitor was measured in MOA 
a factor of 6 brighter than in OGLE. Unless one of the nearby 
field stars seen in the OGLE images is a source of a strong H$\alpha$
emission, which would be measured in the red MOA filter but not in the $I$
OGLE filter. This possibility is, however, unlikely. Finally, note that
measurements of the BLG-360 progenitor made with the red filter in the MACHO 
project (see Appendix~\ref{macho_sect})
gave values significantly fainter than the $I$ values of OGLE, as expected.
This further strengthens our conclusion that the reference flux given in the
archive MOA data is wrong, i.e. significantly too high.

\section{OGLE-2002-BLG-360 in the archives of the MACHO project}
\label{macho_sect}

\begin{figure}
  \includegraphics[scale=0.4]{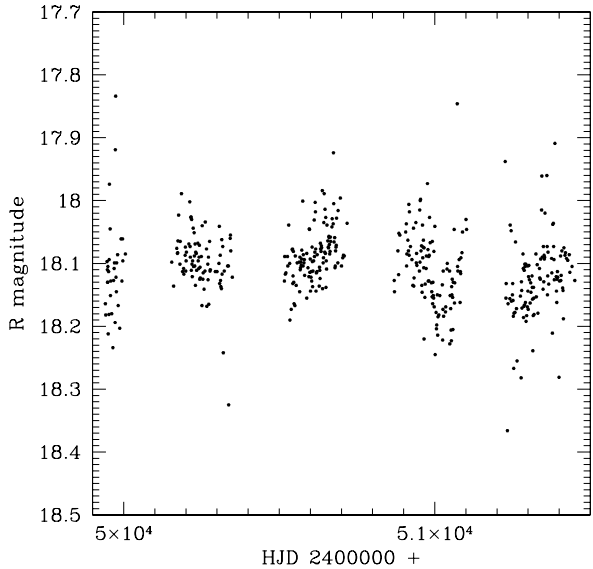}
  \caption{Results of measurements of the BLG-360 progenitor in
1995--99 extracted from the MACHO archives. The observations were made with 
the MACHO red filter.}
  \label{macho_r_fig}
\end{figure}

A field including the position of OGLE-2002-BLG-360 was monitored in the
MACHO project \citep{macho} in 1995-99. 
The observations were obtained with the blue and
red filters of MACHO. Measurements from the blue filter images show a large
scatter in the range of 19--20 magnitudes and are most probably strongly
affected by
scattered light from nearby field stars. We therefore analyse the results
of the red filter observations alone, which are shown in Fig.~\ref{macho_r_fig}.

\begin{figure}
  \includegraphics[scale=0.4]{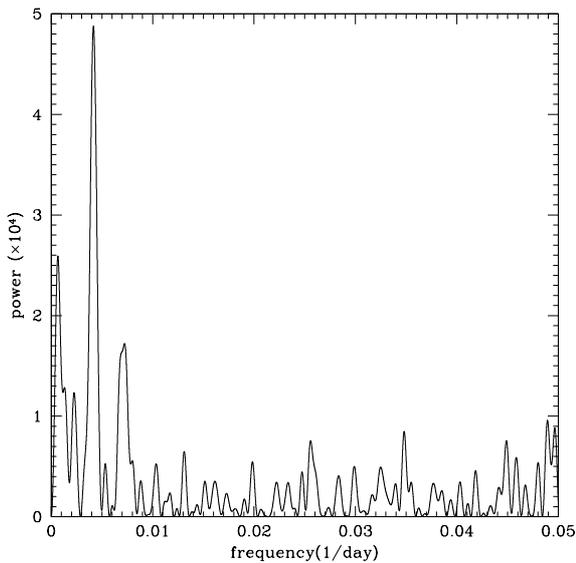}
  \caption{Power spectrum of photometric variations of the BLG-360 progenitor in
1995--99 derived from the MACHO red filter measurements.}
  \label{macho_ps_fig}
\end{figure}

\begin{figure}
  \includegraphics[scale=0.4]{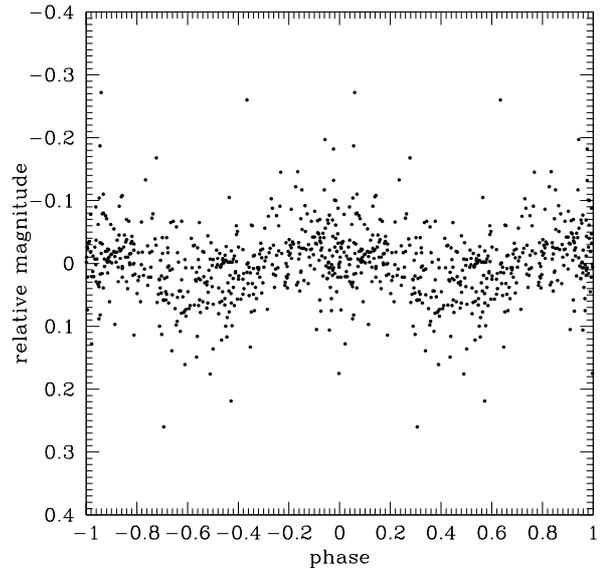}
  \caption{Photometric mesurements of the progenitor of BLG-360 in
1995--99 folded with a period of 240~days, which corresponds to the highest peak
in the power spectrum presented in Fig.~\ref{macho_ps_fig}.}
  \label{macho_lc_fig}
\end{figure}

As can be see from Fig.~\ref{macho_r_fig}, five years of monitoring the
progenitor of BLG-360 do not show any significant long term evolution of the
object. The results oscillate around a mean value of $\sim$18.1 mag. Most of
this fluctuations are probably of instrumental origin and contamination from
light of nearby field stars. One can, however, notice
$\sim$0.1~mag variations of the object on a time scale of months. We
performed a periodogram analysis of the data. A power spectrum of
the variations is shown in Fig.~\ref{macho_ps_fig}. 
As can be seen from the figure,
the most significant peak in the power spectrum corresponds to a
period of $\sim$240~days (for frequences
higher than those shown in the figure the power spectrum shows only peaks
corresponding to one day and its harmonics). The obtained power of this peak
corresponds to an amplitude of 0.022 mag. The observational data folded with
a period of 240~days are shown in Fig.~\ref{macho_lc_fig}.

\end{appendix}

\end{document}